\documentclass[aps,pre,twocolumn,superscriptaddress,showpacs]{revtex4}
\usepackage{amssymb}
\usepackage{epsfig}
\usepackage{amsmath}
\usepackage{times}

\begin{document}

\title{Enhancing synchronization in complex networks of coupled phase oscillators}

\author{Xingang Wang}

\affiliation{Temasek Laboratories, National University of
Singapore, Singapore, 117508}

\affiliation{Beijing-Hong Kong-Singapore Joint Centre for
Nonlinear \& Complex Systems (Singapore), National University of
Singapore, Kent Ridge, Singapore, 119260}

\author{Shuguang Guan}

\affiliation{Temasek Laboratories, National University of
Singapore, Singapore, 117508} \affiliation{Beijing-Hong
Kong-Singapore Joint Centre for Nonlinear \& Complex Systems
(Singapore), National University of Singapore, Kent Ridge,
Singapore, 119260}

\author{Ying-Cheng Lai}

\affiliation{Department of Electrical Engineering, Department of
Physics and Astronomy, Arizona State University, Tempe, Arizona
85287, USA}



\author{Choy Heng Lai}

\affiliation{Beijing-Hong Kong-Singapore Joint Centre for
Nonlinear \& Complex Systems (Singapore), National University of
Singapore, Kent Ridge, Singapore, 119260} \affiliation{Department
of Physics, National University of Singapore, Singapore, 117542}

\begin{abstract}
By a model of coupled phase oscillators, we show analytically how
synchronization in {\em non-identical} complex networks can be
enhanced by introducing a proper gradient into the couplings. It
is found that, by pointing the gradient from the large-degree to
the small-degree nodes on each link, increase of the gradient
strength will bring forward the {\em onset} of network
synchronization monotonically, and, with the same gradient
strength, heterogeneous networks are more synchronizable than
homogeneous networks. The findings are tested by extensive
simulations and good agreement are found.
\end{abstract}

\date{\today }

\maketitle

Synchronization in complex networks has been a topic of arising
interest in recent years, mainly due to its implications to the
practical processes observed in biological and neural systems
\cite{NETWORK:REVIEW,NETSYN:REVIEW}. While most of the studies are
focusing on the phenomenon of complete synchronization in networks
of identical node dynamics \cite{NETWORK:IDENTICAL}, there are
also interests in the collective behaviors in non-identical
networks \cite{ROH:PRL,ROH:PRECHAOS}. The model of non-identical
network is more representative to the realistic situations and, to
analyze its dynamical properties, requires some special
mathematical methods \cite{ROH:PRL,ROH:PRECHAOS}. Different to the
studies in identical networks, in non-identical networks people
are usually interested in the onset of the system coherence, i.e.
the critical coupling from where the systems transits from the
incoherent to coherent states \cite{ROH:PRECHAOS,ONSET:OTHERS}.
For general node dynamics, the critical coupling can be estimated
by the method presented in Ref. \cite{ROH:PRL}, based on the
information of the node dynamics and the largest eigenvalue of the
adjacency matrix of the network. However, for the special case of
coupled phase oscillators, the onset of network synchronization
could be described more accurately by some other approaches. For
instance, it has been shown that the critical coupling
characterizing the onset can be efficiently predicted based on
only the information of network degree distribution, i.e. the
mean-field (MF) approach in Ref. \cite{ROH:PRECHAOS}. In all these
studies, the network couplings are considered as of uniform
strength, i.e. the unweighted networks.

Noticing that couplings in realistic networks are usually directed
and weighted and, in many cases, the direction and weight of the
couplings are determined by a scalar field \cite{TB:2004}, it is
thus natural to extend the study of non-identical network to the
weighted case. For identical networks, it is shown that the
synchronizability of a complex network can be significantly
improved by introducing gradient into the couplings
\cite{HCAB:2005,NETWORK:GRADIENT,WLL:2007}. So far these findings
are obtained from {\em identical networks} and referring to the
transition of {\em global network synchronization}. In this paper,
we are going to study the effects of coupling gradient on the {\em
onset of system coherence in non-identical networks}, and explore
their dependence to the network topology. The former study will
extend the MF approach of Ref. \cite{ROH:PRECHAOS} to the
situation of weighted network and the later stay could provide
theoretical support to the findings of synchronization pathes in
Ref. \cite{GMA:2007}.

We consider network of $N$ coupled phase oscillators of the
following form (the generalized Kuramoto model
\cite{ROH:PRECHAOS})
\begin{equation}
\dot{\theta}_{n}=\omega _{n}+\varepsilon
\underset{m=1}{\overset{N}{\sum}}C_{nm}\sin (\theta _{m}-\theta
_{n}), \label{Kuramoto}
\end{equation}
with $\theta_{n}$ and $\omega_{n}$ the phase and natural frequency
of oscillator $n$ respectively, $\varepsilon$ is the overall
coupling strength, and $C_{nm}$ is an element of the coupling
matrix $C$. In general, the matrix $C$ is asymmetrical and the
frequency $\omega_{n}$ follows some probability distribution
$\rho(\omega)$. For the purpose of theoretical tractability, we
assume that the network is densely connected and has a large size.
Defining the global order parameter as $r\equiv
\sum_{n=1}^{N}r_{n}/\sum_{n=1}^{N}d^{in}_{n}$, with $r_{n}e^{i\psi
_{n}}\equiv \sum_{m=1}^{N}C_{nm}\left\langle e^{i\theta
_{m}}\right\rangle _{t}$ the local order parameter and
$d^{in}_{n}\equiv\sum_{m=1}^{N}C_{nm}$ the total {\em incoming}
couplings of $n$, then the onset of the network synchronization is
characterized by the critical coupling strength $\varepsilon_{c}$
at which $r$ starts to increase from $0$. By the approaches of
Ref. \cite{ROH:PRECHAOS}, we are able to obtain a similar equation
for $r$ in the region of $\varepsilon \geq \varepsilon_{c}$
\begin{equation}
r^{2}= \frac{1}{\alpha _{1}\alpha _{2}^{2}} \frac {\left\langle
d^{in}d^{out}\right\rangle ^{3}}{ \left\langle \left(
d^{in}\right) ^{3}d^{out}\right\rangle \left\langle
d^{in}\right\rangle ^{2}} \left( \frac{ \varepsilon }{\varepsilon
_{c}}-1\right) \left( \frac{\varepsilon }{ \varepsilon
_{c}}\right)^{-3}, \label{MF}
\end{equation}
with $d^{out}_{n}\equiv\sum_{m=1}^{N}C_{mn}$ is the total outgoing
couplings departing from $n$, $\alpha_{1}=2/[\pi g(0)]$ and
$\alpha_{2}=-\pi g^{''}(0)\alpha_{1}/16$ are two parameters
determined by the first-order and second-order approximations of
the frequency distribution $\rho(\omega)$, respectively. The
critical coupling is given by the following equation
\begin{equation}
\varepsilon _{c}=\alpha \frac{\left\langle d^{in}\right\rangle }{
\left\langle d^{in}d^{out}\right\rangle },  \label{threshold}
\end{equation}
with $\left\langle \ldots \right\rangle$ denotes the system
average. Please note that in our weighted model, the total
incoming couplings $d^{in}$ and the total outgoing couplings
$d^{out}$ of each node are real and, in general, unequal. Our main
task is to investigate how the distributions of $d^{in}$ and
$d^{out}$ will affect the onset of network synchronization.

We start by considering an unweighted, symmetrical network
described by adjacency $A={a_{nm}}$, with $a_{nm}=1$ if nodes $n$
and $m$ are connected, $a_{nm}=0$ otherwise, and $a_{n,n}=0$. The
degree of node $n$ is $k_{n}=\sum^{N}_{m=1} a_{nm}$. To introduce
gradient into the couplings, we transform matrix $A$ as follows.
For each pair of connected nodes $n$ and $m$ in the network, we
deduce an amount $g$ from $a_{nm}$ (the coupling that $n$ receives
from $m$) and add it to $a_{mn}$ (the coupling that $m$ receives
from $n$). In doing this, the total couplings between $n$ and $m$
is keeping unchanged. Therefore a coupling gradient is generated
which is pointing from node $n$ to node $m$. Denoting the resulted
matrix as $S$. The coupling matrix $C$ is then defined as $C_{nm}
\equiv k_{n}s_{nm}/\sum^N_{j=1}s_{nj}$ for the non-diagonal
elements, and $C_{nn}=k_{n}$ for the diagonal elements. The
coupling gradient from $n$ to $m$ thus is $\Delta
C_{mn}=C_{mn}-C_{nm}=k_{m}s_{mn}/\sum^N_{j=1}s_{mj}-k_{n}s_{nm}/\sum^N_{j=1}s_{nj}$.
In realistic systems, the direction and weight of each gradient
are generally determined by a unified scalar field which, in the
sense of network synchronization, is usually defined on the node
degree \cite{TB:2004,HCAB:2005,WLL:2007}. Without losing
generality, we make the gradient point from larger-degree to
smaller-degree nodes on each link (the inverse case can be
achieved by $g<0$).

Now we discuss how the change of the gradient parameter $g$ will
affect the network synchronization. Noticing that in Eq.
(\ref{threshold}) the value of $\alpha$ is independent of $g$ and,
by the definition of $C$, we always have $\left\langle d^{in}
\right\rangle = \left\langle k \right\rangle$, which is also
independent of $g$. Therefore the introduction of gradient will
only affect the value of $\left\langle d^{in} d^{out}
\right\rangle$. Rearranging the node index by a descending order
of their degrees, i.e. $k_{1}>k_{2} \ldots
>k_{N}$, then the outgoing couplings of $n$ can be
divided into two groups. Neighbors of node index $m<n$ have
element $s_{mn}=1-g$ in matrix $S$ and $C_{mn}<1$ in matrix $C$
(gradient points to $n$), while for nodes of index we have
$s_{mn}=1+g$ in matrix $S$ and $C_{mn}>1$ in matrix $C$ (gradient
points to $m$). By this partition, the total outgoing couplings
$d_{n}^{out}$ is approximated as
\begin{equation}
d_{n}^{out}=k_{n}\left\{ \frac{1-g}{\Omega
_{i}}P_{i<n}+\frac{1+g}{\Omega _{i}}P_{i>n}\right\}, \label{d_out}
\end{equation}
with $P_{i<n}$ ($P_{i>n}$) the probability for a randomly chosen
node to have degree larger (smaller) than node $n$. $\Omega
_{i}=\frac{1}{k_{i}}\sum^{N}_{j=1}s_{ij}$ is the normalizing
factor defined on node. In calculating $\Omega_{i}$, again, we can
divide the neighbors of $i$ into two groups. Nodes of index $j<i$
have $s_{ij}=1+g$ and nodes of index $j>i$ have $s_{i,j}=1-g$.
Based on this partition, we write
\begin{equation}
\Omega _{i}=1+g (P_{j<i}-P_{j>i}),  \label{Omega}
\end{equation}
with $P$ the same definition as that of Eq. (\ref{d_out}). For
heterogeneous networks of degree distribution $P(k)= Ck^{-\gamma}$
and , we have $\Omega _{i}=1+\frac{gC}{\gamma -1}\left\{
2k_{i}^{1-\gamma }-k_{\max }^{1-\gamma }-k_{\min }^{1-\gamma
}\right\}$, with $k_{max}$ and $k_{min}$ denote the largest and
smallest node degrees of the network, respectively. Inserting this
into Eq. (\ref{d_out}), we obtain
\begin{equation}
d_{n}^{out}=k_{n}\left[ F+G_{n}\right]   \label{d_theory}
\end{equation}
with
\begin{equation}
F=\frac{1}{2g }\left\{ \left( 1+g \right) \ln \left( 1+g \right)
-\left( 1-g \right) \ln \left( 1-g \right) \right\} \label{F}
\end{equation}
and
\begin{equation}
G_{n}=-\ln \left[ 1+g \frac{k_{\max }^{1-\gamma }+k_{\min
}^{1-\gamma }-2k_{n}^{1-\gamma }}{k_{\max }^{1-\gamma }-k_{\min
}^{1-\gamma }}\right] . \label{G}
\end{equation}
Finally we have
\begin{equation}
\left\langle d^{in}d^{out}\right\rangle =\int_{k_{\min }}^{k_{\max
}}\left[ F+G(k) \right] k^{2}P\left( k\right) dk  \label{din-dout}
\end{equation}
Eq. (\ref{din-dout}) is our main result which tells how the
network synchronization ($\varepsilon_{c}$) changes with the
coupling gradient ($g$) and the network topology ($\gamma$).

From Eq. (\ref{d_theory}) we know that, in comparison with the
unweighted networks, the introduction of the coupling gradient
changes only the weight $H\equiv F+G$ of the outgoing couplings on
each node, while in this process the total strength of the
outgoing couplings is keeping unchanged. That is to say, gradient
makes the distribution of $H$ change from an even form ($H\equiv
1$ in unweighted network) to an uneven form ($H=H(g,k)$ in
weighted network). Physically, the term $F$ can be understood as a
summation of the symmetrical part of the couplings on each node,
i.e. $F\sim \sum_{i=1}^{N}\min(C_{ni},C_{in})$, which only depends
on parameter $g$ and will be decreased as $g$ is increased. In
contrast, the term $G$ is a joint function of $g$ and $k_{n}$.
While $G$ increases with $g$, its exact value, however, are
strongly modified by the node degree: larger degree assumes larger
$G$ [Eq. (\ref{G})]. The joint effect of $F$ and $G$ will divide
the nodes into two groups. Nodes of degrees larger than some
critical value $k_{c}$ have weight $H>1$, while nodes of degrees
smaller than $k_{c}$ have weight $H>1$. The critical degree
$k_{c}$ can be calculated from the equation of $H=1$. Under the
assumption of $k_{max}>>k_{min}$, we have
\begin{equation}
k_{c}=\ln \left[ \frac{1}{2}-\frac{1}{2g }\left( 1-e^{F-1}\right)
\right] ^{\frac{1}{1-\gamma }}k_{\min }. \label{degree_threshold}
\end{equation}
The uneven distribution of $H$ can be further understood by
considering its approximations at $k\approx k_{max}$ and $k\approx
k_{min}$, which results in $H_{k\approx
k_{max}}=\frac{1}{2g}\left( 1+g\right)\ln \frac{1+g}{1-g}$ and
$H_{k\approx k_{min}}=\frac{1}{2g}\left( 1-g\right)\ln
\frac{1+g}{1-g}$. Clearly, we have $H_{k\approx k_{max}}>
H_{k\approx k_{min}}$. Since the sum of $H$ over the network is
fixed, i.e. $\sum^{N}_{i=1} H_{i} = N$, the gradient effect thus
can be roughly regarded as a shifting of weight $H$ from
smaller-degree to higher-degree nodes.

For scale-free networks generated by the standard BA growth model
\cite{NETWORK:REVIEW}, we have $k_{max}\approx k_{min}N^{\frac
{1}{\gamma -1}}$. Inserting this relation into Eq. (\ref{G}) we
obtain
\begin{equation}
H=F-\ln \left[ 1-\beta +2\beta \left( \frac{k}{k_{\min }}\right)
^{1-\gamma } \right],   \label{exponent}
\end{equation}
which basically tells the following: for fixed gradient parameter
$g$, increasing the homogeneity of the network, i.e. increasing
exponent $\gamma$, will make the distribution of $H$ more
homogeneous and, as a result, the network synchronization will be
suppressed (i.e. the value of $\varepsilon_{c}$ will increase with
$\gamma$). Eq. (\ref{exponent}) gives the dependence of network
synchronization on network topology.

Now the effect of coupling gradient and the effect of topology on
the starting of synchronization in nonidentical networks can be
summarized as follows. The changes of the gradient strength $g$ or
the degree exponent $\gamma$ do not change the total coupling cost
of the network, they will only redistribute the weights of the
outgoing couplings at each node according to its degree
information. By adding gradient, the outgoing couplings at the
small-degree nodes (of degree $k<k_{c}$) will be reduced by an
amount and added to those of large-degree nodes (of degree
$k>k_{c}$). This will induce a heterogeneous distribution in $H$
which in turn will decrease the threshold coupling
$\varepsilon_{c}$ (see Eq. (\ref{threshold})). This enhancement of
network synchronization, however, is modulated by the network
topology. By increasing the degree exponent $\gamma$, the
distribution of $H$ tends to be homogeneous (i.e. $H \sim 1$) and,
consequently, network synchronization is suppressed. These are the
mechanisms governing the effects of gradient and topology on
network synchronization. The above analysis shows that: 1) the
synchronization of non-identical networks can be enhanced by
coupling gradient; and 2) in comparison with homogeneous networks,
heterogeneous networks take more advantages from the coupling
gradient.

\begin{figure}[tbp]
\begin{center}
\epsfig{figure=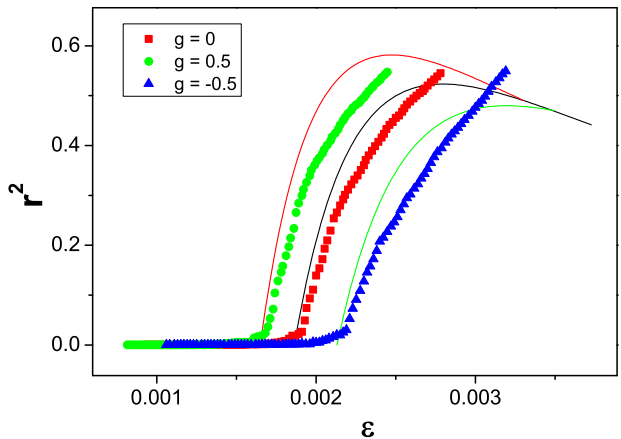,width=\linewidth} \caption{(Color
online) For scale-free network of 1500 nodes, average degree 400,
and degree exponent $\gamma =3$, the variation of the squared
order parameter $r^{2}$ as a function of the coupling strength
$\varepsilon$ in the region of $\varepsilon \in \left [
0.5\varepsilon_{c}, 1.5\varepsilon_{c} \right ]$ by using gradient
parameters $g=0.5$ (the left symbol curve), $g=0$ (the middle
symbol curve), and $g=-0.5$ (the right symbol curve). Apparently,
the onset point of synchronization is shifted to the small values
as $g$ increases. Each data is averaged over $10$ network
realizations. The three line curves are plotted according to Eq.
(\ref{MF}), which predicts the behavior of $r^{2}$ reasonably well
in the region of $\varepsilon \in \left [ \varepsilon_{c},
0.3\varepsilon_{c}\right ]$. In all the three cases, the numerical
results of the critical couplings $\varepsilon_{c}$ are in good
agreements with the theoretical predictions calculated from Eq.
(\ref{threshold}).} \label{fig:MFApproach}
\end{center}
\end{figure}

We now provide the numerical results. The networks are generated
by a generalized BA model \cite{BA:GENERAL}, which is able to
generate networks of varying degree exponent $\gamma$. The
frequency distribution is given by $\rho(\omega)=(3/4)(1-\omega
^{2})$ for $-1<\omega <1$ and $\rho(\omega)=0$ otherwise. The
initial phase $\theta$ of each oscillator is randomly chosen
within range $\left [ 0, 2\pi \right ]$. A transition time $T=100$
is discarded, and the value of $r^{2}$ is calculated over another
period of $T=100$. To show the gradient effects on network
synchronization, we have calculated the variations of the squared
order parameter $r^{2}$ as a function of the coupling strength
$\varepsilon$ for three different gradient parameters: $g=0$,
$0.5$ and $-0.5$. (According to our definition, $g<0$ means that
gradient is pointing from smaller to larger nodes.) The results
are plotted in Fig. \ref{fig:MFApproach}. Clearly, the critical
coupling strength $\varepsilon_{c}$ is shifted to small values as
$g$ is increased. The three lines plotted in Fig.
\ref{fig:MFApproach} represents the theoretical results of Eq.
(\ref{MF}), which fit well with the numerical results in the
neighboring region of the onset. More importantly, the position of
the onset coupling $\varepsilon_{c}$ is predicted precisely by Eq.
(\ref{threshold}). (The precision of this predication is dependent
on the size and connectivity of the network, larger and denser
networks give better results.)

\begin{figure}[tbp]
\begin{center}
\epsfig{figure=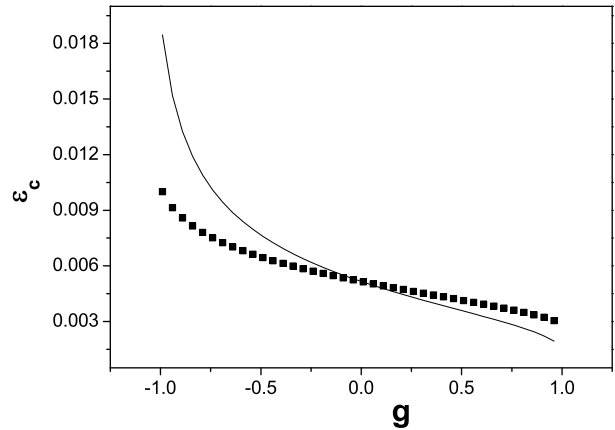,width=\linewidth} \caption{For a
scale-free network of 5000 nodes, average degree 100, and degree
exponent $\gamma =3$, the variation of the critical coupling
strength $\varepsilon_{c}$ as a function of the gradient parameter
$g$. The solid line represents the theoretical results predicted
by Eq. (\ref{d_out}).} \label{fig:GradientEffect}
\end{center}
\end{figure}

To have a global picture on the gradient effect, we plot Fig.
\ref{fig:GradientEffect} the simulation result of the variation of
$\varepsilon_{c}$ as a function of $g$. It is shown that, as $g$
changes from $-1$ to $1$, the value of $\varepsilon_{c}$ is {\em
monotonically} decreased. This process of synchronization
enhancement is well captured by Eq. (\ref{G}), especially in the
region of $g>0$. Since in our analysis we have assumed the network
to be of very large size and of dense connectivity, the mismatch
between the theoretical and numerical results in Fig.
\ref{fig:GradientEffect} is reasonable.

\begin{figure}[tbp]
\begin{center}
\epsfig{figure=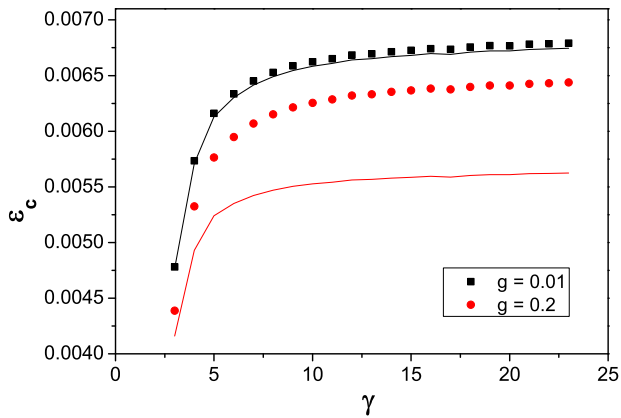,width=\linewidth} \caption{(Color
online) For a scale-free network of 5000 nodes, average degree
100, the variation of the critical coupling strength
$\varepsilon_{c}$ as a function of the degree exponent $\gamma$
under the gradient parameters $g=1\times 10^{-2}$ (the upper
symbol curve) and $g=0.2$ (the lower symbol curve). For both
cases, $\varepsilon_{c}$ increases with $\gamma$. The solid lines
are the theoretical results predicted by Eq. (\ref{exponent}).}
\label{fig:ExponentEffect}
\end{center}
\end{figure}

Simulations have been also conducted on the dependence of
$\varepsilon_{c}$ on $\gamma$. By the generalized BA model
\cite{BA:GENERAL}, we vary the degree exponent $\gamma$
continuously from $3$ to $25$, while keeping the size and average
degree of the network unchanged. As we have predicted [Eq.
(\ref{exponent})], in Fig. \ref{fig:ExponentEffect} it is found
that, for each value of $g$, the critical coupling
$\varepsilon_{c}$ will increase monotonically with the degree
exponent $\gamma$. Specially, for the case of $g=1\times 10^{-2}$
in Fig. \ref{fig:ExponentEffect}, the numerical results are in
good agreements with the theoretical results of Eq.
(\ref{degree_threshold}). As $g$ increases the mismatch between
the theoretical and numerical results is enlarged, especially for
networks of larger $\gamma$. Again, by increasing size and
coupling density of the network, the mismatch can be alleviated.

A few remarks are in order. Firstly, while our theory gives well
approximations on the collective behavior in densely connected
large networks, our findings about gradient effects of their
dependence to network topologies are general for any network. The
amazing thing is that, for network of given degree distribution
(not limited to the scale-free type), our theory tells how much
improvement could the network benefits from a given gradient. From
the findings, we are able to not only point out clearly the
optimal configuration for synchronization, which happens when
$g=1$ [Fig. \ref{fig:GradientEffect}], but also have a systematic
understanding on the {\em transition} from unweighted to optimal
network, and, more importantly, the underlying mechanisms that
govern this transition. Secondly, although similar findings about
the gradient effects had been discovered previously in the study
of global synchronization of identical networks
\cite{HCAB:2005,NETWORK:GRADIENT,WLL:2007}, our analysises,
however, are focusing on the {\em onset synchronization in
non-identical networks}. Another difference is, by adopting the
generalized Kuramoto model, that we are able to show {\em
analytically} how the coupling gradient affects synchronization
(see Eq. (\ref{din-dout}) and Fig. \ref{fig:GradientEffect}) and
what is the role of network topology in this process ((see Eq.
(\ref{exponent}) and Fig. \ref{fig:ExponentEffect}). It is noticed
that in Ref. [\cite{GMA:2007}] the authors found numerically that
the onset of synchronization in scale-free networks happens in
advance to that of homogeneous networks, which, according to the
approximation of Eq. (\ref{exponent}), can be easily understood.

In summary, we have studied the effects of coupling gradient on
the onset of synchronization in nonidentical complex networks. It
is found that: 1) network synchronization can be enhanced by
introducing gradient into to the couplings; and 2) in terms of the
onset of synchronization, heterogeneous networks are more
synchronizable than homogeneous networks. We hope these findings
to be helpful in understanding the collective behaviors in
realistic systems.

YCL was also supported by AFOSR under Grants No. FA9550-06-1-0024
and No. FA9550-07-1-0045.

\end{document}